\documentclass[%
 reprint,
superscriptaddress,
 amsmath,amssymb,siunitx
 aps,
 prl,
]{revtex4-2}

\usepackage{graphicx}
\usepackage{dcolumn}
\usepackage{bm}
\usepackage{hyperref}
\hypersetup{
  colorlinks   = true, 
  urlcolor     = black, 
  linkcolor    = blue, 
  citecolor   = blue 
}

\newcommand{\ket}[1]{\big| #1 \big>} 
\newcommand{\matrixel}[3]{\big< #1 \vphantom{#2#3} \big| #2 \big| #3 \vphantom{#1#2} \big>} 
\newcommand{\avg}[1]{\left<#1\right>} 
\newcommand{\lr}[1]{\left(#1\right)}

\usepackage{xcolor}

\begin{document}

\preprint{APS/123-QED}

\title{Terminable Transitions in a Topological Fermionic Ladder}

\date{\today}
\author{Yuchi He}
\affiliation{Institut f{\"u}r Theorie der Statistischen Physik, RWTH Aachen University and
  JARA---Fundamentals of Future Information Technology, 52056 Aachen, Germany}
\affiliation{Rudolf Peierls Centre for Theoretical Physics, Clarendon Laboratory, Parks Road, Oxford OX1 3PU, United Kingdom}
\author{Dante M.~Kennes}
\affiliation{Institut f{\"u}r Theorie der Statistischen Physik, RWTH Aachen University and
  JARA---Fundamentals of Future Information Technology, 52056 Aachen, Germany}
 \affiliation{Max Planck Institute for the Structure and Dynamics of Matter, Center for Free Electron Laser Science, 22761 Hamburg, Germany}
 \author{Christoph Karrasch}
\affiliation{Technische Universität Braunschweig, Institut für Mathematische Physik,
Mendelssohnstraße 3, 38106 Braunschweig, Germany}

\author{Roman Rausch}
\affiliation{Technische Universität Braunschweig, Institut für Mathematische Physik,
Mendelssohnstraße 3, 38106 Braunschweig, Germany}

\begin{abstract}
Interacting fermionic ladders are versatile platforms to study quantum phases of matter, such as different types of Mott insulators. In particular, there are D-Mott and S-Mott states that hold preformed fermion pairs and become paired-fermion liquids upon doping ($d$ wave and $s$ wave, respectively).

We show that the D-Mott and S-Mott phases are in fact two facets of the same topological phase and that the transition between them is terminable.  
These results provide a quantum analog of the well-known terminable liquid-to-gas transition. However, the phenomenology we uncover is even richer, as the order of the transition may alternate between continuous and first-order, depending on the interaction details. Most importantly, the terminable transition is robust in the sense that it is guaranteed to appear for weak, but arbitrary couplings.

We discuss a minimal model where some analytical insights can be obtained, a generic model where the effect still persists; and a model-independent field-theoretical study demonstrating the general phenomenon. The role of symmetry and the edge states is briefly discussed. The numerical results are obtained using the variational uniform matrix-product state formalism for infinite systems, as well as the density-matrix renormalization group algorithm for finite systems.
\end{abstract}

\maketitle

A ladder geometry can be thought of as a narrow strip of a two-dimensional lattice, or as a chain endowed with additional local degrees of freedom (the ``rungs" of the ladder).
Ladders that host interacting fermions are versatile flagship platforms for studying quantum phases and their transitions in one dimension~\cite{Giamarchi_book,DagottoRice,BF,SO8, Wu,TsuchiizuFurusaki}, such as repulsion-induced pairing~\cite{PhysRevB.45.5744,Noack1994,Noack_1995,Noack1997,Chakravarty2001,ladderRG,PDW,hirthe2022}; or serve as realizations of symmetry-protected topological phases~\cite{Sompet2022, AnfusoRosch, MoudgalyaPollmann2015, White1996, VMP, SSH, Haldanechain}.
Ladder models also appear for two-orbital chains~\cite{hundchain,sunladder} and effectively for more general quasi-one-dimensional systems, such as nanoribbons~\cite{Mishra2021} and nanotubes~\cite{BF,PhysRevB.78.035401,Varsano2017,CNTmodel,Okamoto2018}.

\begin{figure}[t!]
\begin{center}
\includegraphics[width=0.7\columnwidth]{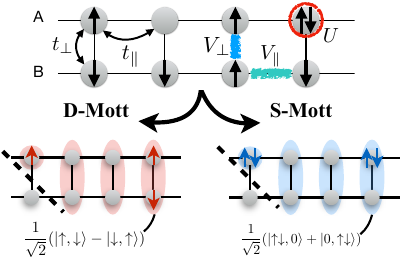}
\end{center}
\caption{
Top: Illustration of our minimal ladder model, Eq.~\eqref{H}.
Bottom: The idealized wave functions of the D-Mott (S-Mott) phase are given by product states of rung singlets (on-site singlets) in the limit of strong local repulsion $U>0$ (strong local attraction $U<0$). For open boundary conditions that cut the singlets open (dotted line), edge states are produced that have spin (charge) degrees of freedom.
The S-Mott state can be equally achieved by a strong intrarung repulsion $V_{\perp}>0$.
}\label{fig:ladder}
\end{figure}

A particularly interesting aspect is that fermionic ladders realize Anderson's mechanism for superconductivity from repulsive interactions, which was originally proposed for cuprates~\cite{Anderson1987,Chakravarty2001}: An effective exchange interaction at half filling causes fermions to pair up as spin singlets in an insulating Mott phase; these preformed pairs become mobile upon doping. While the physics of cuprates has turned out to be more complicated, the finite extension of the rungs of a ladder strongly favors such a pairing with a particularly strong binding energy~\cite{Chakravarty2001}.
Two pairing patterns can occur on a rung (see Fig.~\ref{fig:ladder}): 
If local repulsion dominates, it avoids double occupancy and promotes singlets across the rung. If local attraction dominates, it favors double occupancy and promotes on-site singlets. Upon doping, these patterns yield superconducting states that have been dubbed
``$d$ wave" and ``$s$ wave", respectively, in analogy to the 2D case~\cite{Giamarchi_book,TsuchiizuFurusaki}. The half-filled insulating states are correspondingly called ``D-Mott'' and ``S-Mott''~\cite{SO8,TsuchiizuFurusaki}. It is known that the rung-singlet wave function (D-Mott) is a realization of the topological Haldane phase~\cite{Sompet2022, hundchain, VMP, White1996}.

\begin{figure}[t!]
\begin{center}
\includegraphics[width=\columnwidth]{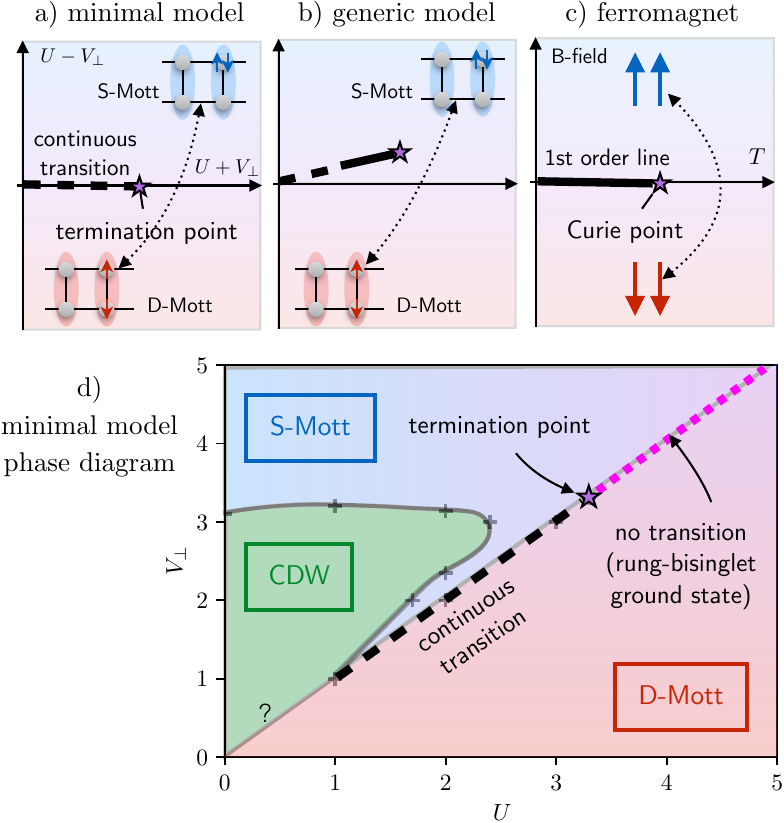}
\end{center}
\caption{(a), (b)
Terminable transitions (schematic) of the fermionic ladder.
The D-Mott and S-Mott  (cf. Fig.~\ref{fig:ladder}) can be adiabatically connected via a path that avoids the transition line. The transition can be tuned to be partially first order for the generic model.
(c)
Paradigm of a terminable transition:
a ferromagnet with a $B$ field at finite temperature $T$. 
Below the Curie point, there is a first-order transition when tuning $B$ across zero, but no transition above it.
The phases are characterized by (a), (b) density difference of D-type and S-type singlets [Eq.~\eqref{orders}]; (c) density difference of $\uparrow$ and $\downarrow$ spins, i.e., magnetic moments.
(d) Quantitative phase diagram for the model Eq.~\eqref{H} for the minimal model, computed by VUMPS (CDW: charge density wave). For small interactions, it is unclear if there is a direct transition between D-Mott and CDW marked by ``?". The continuous transition terminates at $U=V_{\perp}\approx 3.4$, after which the gapped exact rung-bisinglet (see text) is the ground state (magenta line). Different scenarios of the transition for the generic model are shown in Fig.~\ref{fig:Vpara}.
}\label{fig:PDsketch}
\end{figure}
 
In this work, we study the competition between the two singlet types in more detail and find that a phase transition emerges between the two Mott states, \textit{but the transition line is terminable}. Therefore, D-Mott and S-Mott are adiabatically connected, and one should think of them as \textit{two facets of one and the same topological phase}. This physics also provides a quantum analog of the prototypical, classical liquid-to-gas transition, which is terminable and of first order (another example is the ferromagnet, cf. Fig.~\ref{fig:PDsketch}). However, we show that the terminable transition in our system is robust in the sense that it is guaranteed to appear at weak, but arbitrary interactions. Furthermore, its order can change from first order to second, depending on the interaction details. This is schematically summarized in Fig.~\ref{fig:PDsketch}.
As the effective theory of liquid-to-gas transitions~\cite{chaikin_lubensky} was integral to understanding the physics of a wide range of very different systems~\cite{QCDCEP,lltransition,mit,schuler2019thermodynamics,Jimenez2021,sushchyev2022thermodynamics,quantumHallCDW},
understanding the robustness and change of order of the transition line might take a similarly pronounced role.

\paragraph{Hamiltonian.—} As a minimal model to observe the phenomenon, we consider the following Hamiltonian of fermions on a ladder (pictorially shown in Fig.~\ref{fig:ladder}):
\begin{equation}\label{H}
H_{\text{min}} = H_0 + H_{\text{Hub}} + H_{\text{ext}}\\
\end{equation}
with
\begin{align}
H_0=&-t_{\parallel}\sum_{j,l,\sigma}c^{\dagger}_{j+1,l,\sigma}c_{j,l,\sigma}
-t_{\perp}\sum_{j,\sigma}c^{\dagger}_{j,\mathrm{A},\sigma}c_{j,\mathrm{B},\sigma}+h.c.\nonumber \\
H_{\text{Hub}}=&\frac{U}{2}\sum_{j,l} \Delta n_{j,l}\Delta n_{j,l}; ~~ H_{\text{ext}}=V_{\perp}\sum_{j}\Delta n_{j,\mathrm{A}} \Delta n_{j,\mathrm{B}}, \nonumber \\
\end{align}
where $c_{j,l,\sigma}$ ($c^{\dagger}_{j,l,\sigma}$) annihilates (creates) a fermion with spin $\sigma$ at the site $j$ of the leg $l=\mathrm{A},\mathrm{B}$ of the ladder; $\Delta n_{j,l}= n_{j,l}-1= \sum_{\sigma}c^{\dagger}_{j,l,\sigma}c_{j,l,\sigma}-1$ is the density deviation from half filling.

The parameters are as follows: $t_{\parallel}$ ($t_{\perp}$) is the hopping amplitude along the legs (rungs) of the ladder; similarly $V_{\perp}$ is the nearest-neighbor Coulomb interaction along the rungs; $U$ is the local Coulomb interaction. We set $t_{\perp}=t_{\parallel}=1$ and look at a repulsive $U>0$. While local pairing in the S-Mott phase is commonly discussed in the attractive case $U<0$, it can also be achieved by setting $V_{\perp}>0$~\cite{TsuchiizuFurusaki} (see Fig.~\ref{fig:ladder}). Doing so allows us to study the competition between the two pairing patterns in the $U$--$V_{\perp}$ phase diagram without switching off the interaction.

The minimal model allows us to understand the physics most clearly. For ultracold atoms in optical lattices, it is similar to the periodic Anderson model and we believe that both can be realized with equal effort~\cite{Zhong2017}. On the other hand, longer-ranged Coulomb interactions~\cite{Varsano2017} and Hund's rule spin exchange $J_{\perp}<0$~\cite{Omori2011,Tsuchiizu2012} are relevant in materials. To this end, we also study a generic Hamiltonian given by
\begin{equation}\label{Hgen}
H_{\text{gen}} = H_0 + H_{\text{Hub}} + H_{\text{ext}} + H'_{\text{ext}}, \\
\end{equation}
with
\begin{align}\label{Hext_prime}
H'_{\text{ext}}=& V_{\parallel}\sum_{j,l} \Delta n_{j,l}\Delta n_{j+1,l} 
+V'_{\perp} \sum_{j}  (\Delta n_{j,\mathrm{A}}\Delta n_{j+1,\mathrm{B}} + H.c.) \nonumber \\
+ &V'_{\parallel}\sum_{j,l}  \Delta n_{j,l}\Delta n_{j+2,l}
+ J_{\perp}\sum_l \mathbf{S}_{j,\mathrm{A}}\cdot\mathbf{S}_{j,\mathrm{B}},
\end{align}
where $\mathbf{S}_{j,l}$ is the vector of spin operators.
For carbon nanotubes, the parameters $V_{\parallel}\approx V_{\perp}$, $t_{\parallel}\approx t_{\perp}$ are expected to be only slightly anisotropic~\cite{BF,Okamoto2018}. For chemical ladders and two-band systems, they constitute different overlaps and may show stronger anisotropy~\cite{Omori2011,Tsuchiizu2012,DagottoRice,Landee2001,Bisogni2015,Fuseya2007,Pouget2023}. Orbital nematicity may also contribute to anisotropy~\cite{Hosoi2020}.

To solve the model, we employ the \textit{variational uniform matrix product state} (VUMPS) formalism~\cite{White, McCulloch2008, Phien_2012,Haegeman_2016,VUMPS}, which variationally determines the ground state within the class of matrix-product states in the thermodynamic limit. The central control parameter is the ``bond dimension'' $\chi$, which reflects the number of variational parameters. 
This method is able to find ground states of gapped 1D systems to very high accuracy.
We exploit the spin-SU(2) and charge-U(1) symmetry of the underlying problem~\cite{McCulloch_2007}, which allows us to reach bond dimensions of up to $\chi\sim10^4$ in difficult small-gap regions. To look at edge states, we employ the related density-matrix renormalization group (DMRG) algorithm for finite systems~\cite{Schollwoeck}.

Various aspects of this model family have been studied in different parameter regimes. For $V_{\perp}=0$, the main focus has been on the d-wave pairing~\cite{PhysRevB.45.5744,Noack1994,Noack_1995,Noack1997,Karakonstantakis2011,Dolfi2015,ladderRG,hirthe2022,PhysRevB.108.165113,PhysRevB.108.195136}, but also on the excitations~\cite{Weihong_2001,YangFeiguin2019} and the topological properties~\cite{Sompet2022,AnfusoRosch,MoudgalyaPollmann2015,White1996}. For $V_{\perp}\neq0$ and $V_{\parallel}\neq0$, the onset of charge order was studied~\cite{Vojta1999,Robinson2012}.
With analytical methods, phase diagrams have been proposed for various parameter ranges~\cite{BF,SO8,Wu,TsuchiizuFurusaki,Orignac2003,Tsuchiizu2004}. However, the termination of the D-Mott/S-Mott transition and the physics surrounding it have not been revealed in these works.

\begin{figure}[t!]
\begin{center}
\includegraphics[width=\columnwidth]{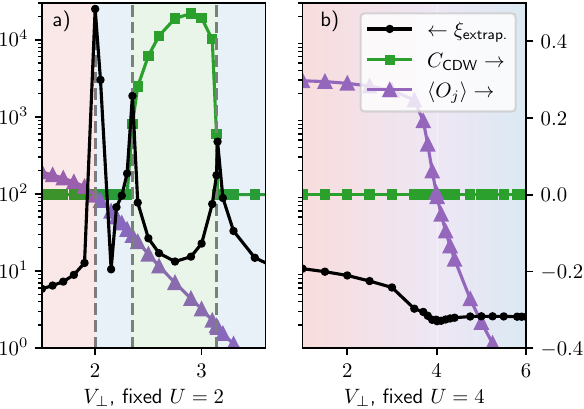}
\end{center}
\caption{
\label{fig:xi}Extrapolated correlation lengths (left scale) and order parameters (right scale) along a) $U=2$; b) $U=4$ for the minimal model Eq.~\eqref{H}. Charge density wave order parameter: $C_{\text{CDW}}=1/2\big|\avg{n_{j,\mathrm{A}}}-\avg{n_{j,\mathrm{B}}}\big|$. For the definition of $\avg{O_j}$, see Eq.~\eqref{orders}.
}
\end{figure}
\paragraph{Microscopic characterization of S-Mott and D-Mott—} 
To characterize S-Mott and D-Mott, we introduce a microscopic order parameter, namely, the ``singlet density difference" $\langle O_j \rangle$:
\begin{align}\label{orders}
&O_j = n_{\mathrm{D}j}-n_{\mathrm{S}j} = \Delta^{\dagger}_{\mathrm{D}j}\Delta_{\mathrm{D}j}-\Delta^{\dagger}_{\mathrm{S}j}\Delta_{\mathrm{S}j}, \nonumber \\
& \Delta_{\mathrm{D}j}=(c_{j,\mathrm{A},\uparrow} c_{j,\mathrm{B},\downarrow}+c_{j,\mathrm{B},\uparrow} c_{j,\mathrm{A},\downarrow})/\sqrt{2}, \nonumber \\
& \Delta_{\mathrm{S}j}=(c_{j,\mathrm{A},\uparrow} c_{j,\mathrm{A},\downarrow}+c_{j,\mathrm{B},\uparrow} c_{j,\mathrm{B},\downarrow})/\sqrt{2}.  
\end{align}
This is motivated by the picture that D-Mott and S-Mott phases host immobile preformed $d$- and $s$-wave pairs~\cite{SO8}, which are characterized by cross-rung pairing (D) and on-site pairing (S)~\cite{Giamarchi_book} (cf. Fig.~\ref{fig:ladder}).
The corresponding pair annihilation operators are $\Delta_{\mathrm{D}j}$ and $\Delta_{\mathrm{S}j}$. In the strongly coupled limit of independent rungs, the prototype states can be constructed as $|\mathrm{D}\rangle=\prod_{j} \Delta^{\dagger}_{\mathrm{D}j}|\Omega\rangle$ and  $|\mathrm{S}\rangle=\prod_{j} \Delta^{\dagger}_{\mathrm{S}j} |\Omega\rangle$~\cite{TsuchiizuFurusaki}, where $|\Omega\rangle$ is the vacuum state (see Fig.~\ref{fig:ladder}). 
Therefore, $\avg{O_j}>0$ $(<0)$ measures that there are more rung (local) singlets in the admixture of the wave function and we expect a sign change across the phase transition.

\paragraph{Results for the minimal model Eq.~\eqref{H}.—} The full phase diagram obtained numerically is shown in Fig.~\ref{fig:PDsketch}. We find a phase transition line between D-Mott and S-Mott along $U=V_{\perp}$, which remarkably terminates at $U=V_{\perp} \approx 3.4$. The continuous transition is detected [Fig.~\ref{fig:xi} (a)] via a divergence of the correlation length $\xi$ in the thermodynamic limit, extrapolated using VUMPS data~\cite{Rams_2018,suppm}. (A direct computation of the gap for finite systems yields consistent results~\cite{suppm}). The region with the charge density wave (CDW) is irrelevant to our discussion. Our data show no gap closing and no obvious discontinuity for large $U$ (cf. $U=4$ in Fig.~\ref{fig:xi} b), implying that there is an adiabatic path connecting the two Mott phases.

The minimal model has an artifact, namely the accidental conservation of particle number in the subband basis for $U=V_\perp$. This enables us to analytically locate the transition exactly along the $U=V_\perp$ line and track its termination. We will later show that the existence of a continuous transition is robust without the need of an accidental symmetry.
Introducing the transverse subband basis $c_{j,k_y,\sigma}$ as 
$c_{j,0,\sigma}=(c_{j,\mathrm{A},\sigma}+c_{j,\mathrm{B},\sigma})/\sqrt{2}$ and 
$c_{j,\pi,\sigma}=(c_{j,\mathrm{A},\sigma}-c_{j,\mathrm{B},\sigma})/\sqrt{2}$,
where $k_y=0,\pi$ is the transverse momentum, we can rewrite the Hamiltonian Eq.~\eqref{H} in this basis:
\begin{align}\label{Hmvpp0}
H_{\text{min}}=&-t_{\parallel}\sum_{j,k_y,\sigma}(c^{\dagger}_{j,k_y,\sigma}c_{j+1,k_y,\sigma}+H.c.)-t_{\perp}\sum_{j}\left(n_{j,\pi}-n_{j,0}\right)\nonumber \\
&+U/2 \sum_j \left(\Delta n_{j,\pi}+ \Delta n_{j,0} \right)^2-(U-V_{\perp})H_{\mathrm{res}},
\end{align}
where $n_{j,k_y}=\sum_{\sigma}c^{\dagger}_{j,k_y,\sigma}c_{j,k_y,\sigma}$.
The residual term $\propto H_{\text{res}}$ vanishes for $U=V_\perp$, so that $N_{\pi}=\sum_{j}n_{j,\pi}$ and $N_{0}=\sum_{j}n_{j,0}$ become conserved. The Lieb-Schultz-Mattis theorem~\cite{LSM,Tasaki2018} states that for a fractional filling factor, the system must be gapless as long as there is no spontaneous breaking of translational symmetry. Our numerics show that the filling ratios $\avg{n_{j,\pi}}$ and $\avg{n_{j,0}}$ are fractional along $U=V_{\perp}$ below the termination point. Above the termination point, the fillings are integer with $\avg{n_{j,\pi}}=0$ and $\avg{n_{j,0}}=2$, supporting the termination of the phase transition line. 
The data are given in the Supplemental Material~\cite{suppm}. Moreover, we find numerically that the ground state above the termination point is given by a product state of equal-weight superpositions of the two singlet types: $\prod_{j} 1/\sqrt{2}(\Delta^{\dagger}_{\mathrm{S}j}+\Delta^{\dagger}_{\mathrm{D}j})|\Omega\rangle$, which we dub ``rung bisinglet" (cf. Fig.~\ref{fig:PDsketch}). It straightforward to show analytically that this is an exact eigenstate of the minimal model for $U=V_{\perp}$. The rung bisinglet can be taken as a simple reference wave function for both the S-Mott and the D-Mott, similar to how the Affleck-Kennedy-Lieb-Tasaki (AKLT) state~\cite{AKLT} is taken as a simple reference wave function for the Haldane phase of the $S=1$ spin chain.

\begin{figure}
\begin{center}
\includegraphics[width=\columnwidth]{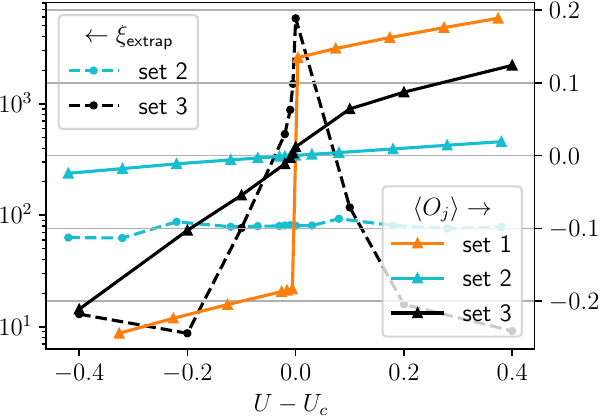}
\end{center}
\caption{Ground-state expectation $\langle O_j \rangle$ (right scale) and extrapolated correlation lengths (left scale) for the generic model Eq.~\eqref{Hgen} and the following datasets:
(1) $t_{\perp}=1.1$, $V_{\perp}=2.88$, $V_{\parallel}=2.4$, $V'_{\perp}=1.92$, $V'_{\parallel}=1.28$, $J_{\perp}=0$ ($U_c\approx 4.426$)
(2) $t_{\perp}=1.1$, $V_{\perp}=0.8$, $V_{\parallel}=0.7$, $V'_{\perp}=0.4$, $V'_{\parallel}=0$, $J_{\perp}=-0.6$ ($U_c\approx 1.52$)
(3) $t_{\perp}=1.1$, $V_{\perp}=2.45$, $V_{\parallel}=0.5$, $V'_{\perp}=0.4$, $V'_{\parallel}=0.3$, $J_{\perp}=0$ ($U_c\approx 2.5$),
exemplifying a first-order transition, no transition and a continuous transition, respectively.
\label{fig:Vpara}
}
\end{figure}
\paragraph{Results for the generic model Eq.~\eqref{Hgen}.—} 
The above accidental symmetry is lifted for generic interactions. The parameter space now becomes quite large and competition from other phases increases. Nevertheless, in Fig.~\ref{fig:Vpara}, we show exemplary cases that illustrate different scenarios: The transition now may become first-order, exhibiting a jump in $\langle O_j \rangle$, but remains continuous for other parameters. We also find instances without a transition.
Thus, numerical evidence indicates that the terminable transition is generic, both for nearly-isotropic and anisotropic interactions.

\paragraph{General effective field theory.—}

We present an effective theory of the S-D Mott transition, which demonstrates (1) that the continuous transition line is robust beyond the accidental symmetry of the minimal model; (2) its potential instability to first order for strong interaction; and (3) its termination.

We use the bosonization approach and ensure that the exact accidental subband symmetry is generically absent. The bosonization of continuum operators corresponding to those of Eq.~\eqref{Hmvpp0} are given by $c_{k_y,\sigma}(x_j)
=
\frac{\kappa_{k_y,\sigma}}{\sqrt{2\pi}} \sum_{\eta=-1,1}e^{i[\theta_{k_y,\sigma}+\eta(\phi_{k_y,\sigma}+k_{\text{F},k_y\sigma}x_j)]}$, 
where $\theta_{k_y,\sigma}(x_j)$ and $\phi_{k_y,\sigma}(x_j)$ are dual to each other satisfying $[\theta_{k_y,\sigma}(x,t), \phi_{k'_y,\sigma'}(x',t)]=i\pi\delta_{k_y,k'_y}\delta_{\sigma,\sigma'}\Theta(x-x')$;
$\{\kappa_{k_y,\sigma},\kappa_{k'_y,\sigma'} \}=2\delta_{k_y,k'_y} \delta_{\sigma,\sigma'}$. The $k_{\mathrm{F},k_y,\sigma}$ is the base wave vector of the low-energy excitation of $c_{j,k_y,\sigma}$.  The half-filling condition fixes $k_{\mathrm{F},0,\sigma}+k_{\mathrm{F},\pi,\sigma}=\pi$, where $k_{\mathrm{F},k_y,\sigma}$ 
is influenced by interaction besides $t_{\parallel}$ and $t_{\perp}$. Two-band bosonization requires partially-filled subbands ($k_{\mathrm{F},k_y,\sigma}\neq0,\pm \pi$).
Introducing a transformed basis for the effective fields:
$ \tilde{\phi}_{c,\pm}=\frac{1}{2}[(\phi_{0,\uparrow}+\phi_{0, \downarrow})\pm(\phi_{\pi,\uparrow}+\phi_{\pi, \downarrow})]$ and
$ \tilde{\phi}_{s,\pm}=\frac{1}{2}[(\phi_{0,\uparrow}-\phi_{0,\downarrow})\pm (\phi_{\pi,\uparrow}-\phi_{\pi,\downarrow})]
$, S-Mott and D-Mott have been defined~\cite{SO8,TsuchiizuFurusaki} as $\tilde{\phi}_{c,+}$, $\tilde{\phi}_{s,+}$, $\tilde{\phi}_{s,-}$ all locked at 0, and $\tilde{\theta}_{c,-}$ locked at $0$ and $\pi/2$ mod $\pi$ respectively. 

We now show the transition line is Gaussian-critical where
$O_j$ has quasi-long-range order and its scaling dimension indicates the instability of Gaussian criticality to a first-order transition when removing the accidental symmetry.
When the sectors other than $(c,-)$ are kept locked, we can approximate the locked fields as constant and obtain
\begin{align}\label{obosonization}
O_{j} \propto -\cos[2\tilde{\theta}_{c,-}(x_j)],    
\end{align}
whose expectation values flip sign when the locking value $\tilde{\theta}_{c,-}=0$ changes to $\pi/2$. The discreteness of locking values is related to time-reversal symmetry, as terms like $\cos[2\tilde{\theta}_{c,-}(x_j)+\alpha]$ with continuous varying $\alpha$ are forbidden by it~\cite{suppm}. Near the Gaussian criticality, the effective Hamiltonian density, neglecting higher harmonics, is
\begin{align}\label{Hc-}
	\mathcal{H}_{c,-} &= \frac{v_{c,-}}{2\pi}\bigg[K(\partial_{x}\tilde{\theta}_{c,-})^2 + \frac{1}{K}(\partial_{x}\tilde{\phi}_{c,-})^2\bigg]+g \cos(2\tilde{\theta}_{c,-}),
\end{align}
where $K$ is the Luttinger parameter and $g \propto (V_{\perp}-U)$ for the minimal model (in the generic case the relation is not known exactly).
Equations~\eqref{obosonization} and~\eqref{Hc-} can be used to predict the correlator $\langle O_j O_{j+d} \rangle \propto 1/|d|^{2/K}$ at the criticality ($g=0$). The scaling dimension of $O_j$ is thus $1/K$. Observing nonuniversal exponents numerically confirms Gaussian criticality.  For example, our minimal-model data~\cite{suppm} suggest that $1/K$ goes down from $\sim 0.96$ to $\sim 0.46$ when increasing $U=V_{\perp}$ from 2 to 3.2. As the 0-loop renormalization group relevance criterion is to have a scaling dimension $<2$, the measured $1/K$ is consistent with this as long as $g \neq 0$ and $\tilde{\theta}_{c,-}$ gets locked.

The stability of the Gaussian transition is controlled by higher harmonic terms like $\cos(4\tilde{\theta}_{c,-})$ ($\sim O^2$); it generically exists in the ``bare" Hamiltonian, contributing to Eq.~\eqref{Hc-} and does not generically vanish simultaneously with  $\cos(2\tilde{\theta}_{c,-})$ unless there is an exact subband U(1) symmetry.
Stable Gaussian criticality requires that those terms are irrelevant, the criterion for which is $1/K>1/2$.  This condition is always satisfied near the weak coupling limit where $1/K \rightarrow 1$, far from the marginal value, so the continuous transition is robust. Depending on model details, a perturbing interaction may induce an instability. For example, we could add some longer-ranged interaction terms as in Eq.~\eqref{Hgen}, and reach $1/K$ that is small enough to induce a first-order transition described by a Landau-Ginzburg theory with powers of $O$, which can describe the transition termination.

\begin{figure}
\begin{center}
\includegraphics[width=\columnwidth]{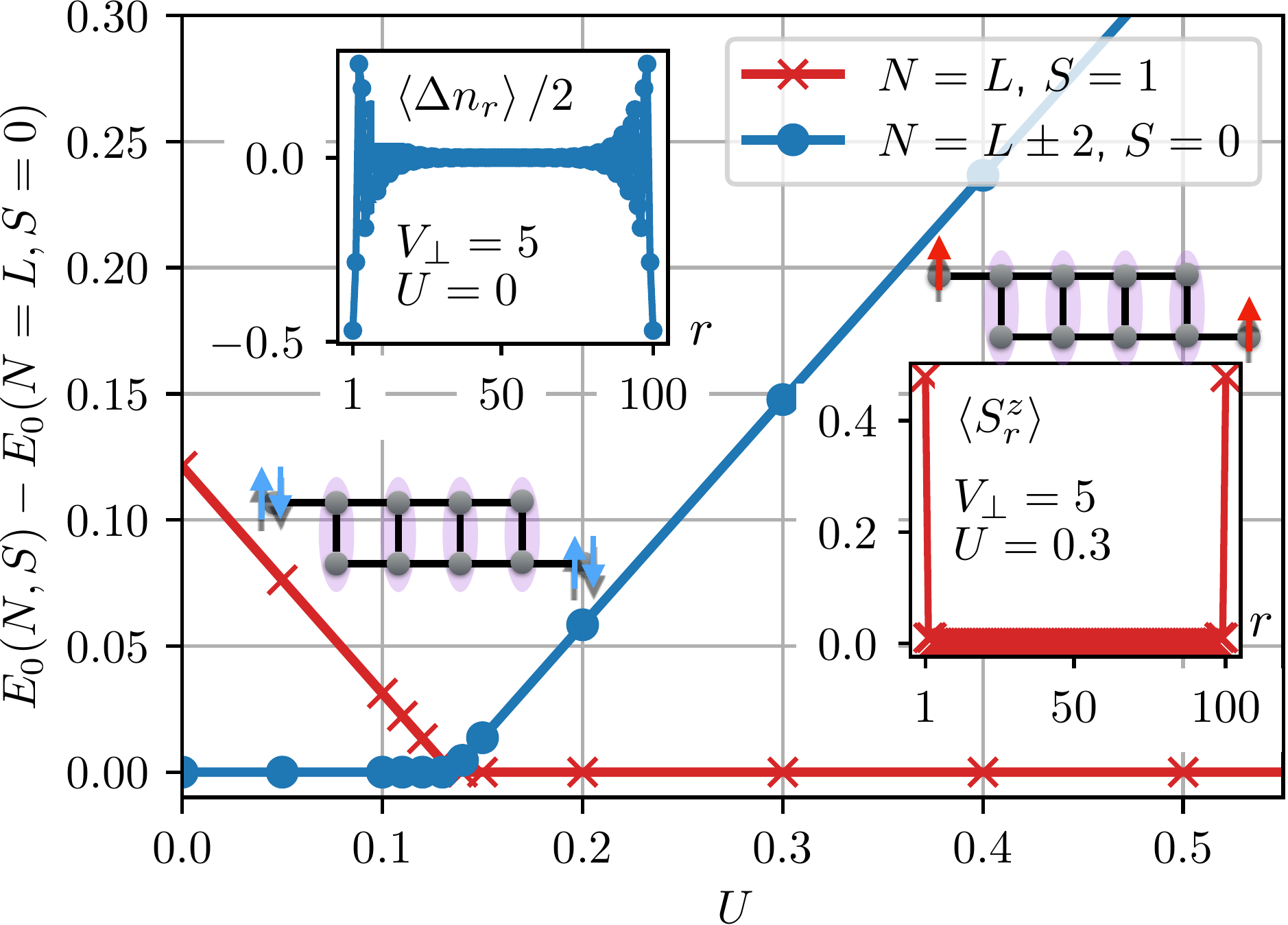}
\end{center}
\caption{
\label{fig:edgeStates}
Ground-state degeneracy and edge modes of open ladders under diagonal cuts. Parameters: number of sites $L=100$ (i.e., $50$ rungs), $V_{\perp}=5$ in the minimal model Eq.~\eqref{H}.
The index $r=2j+l$ ($l=0,1$) consecutively labels the $L$ sites.
$E_0(N,S)$ is the lowest energy with $N$ particles and total spin $S$.
The insets show that the edge modes carry spin and charge quantum numbers, respectively.
Upper left inset: Particle density $1/2\avg{\Delta n_{r}} = 1/2\lr{\sum_{\sigma}\avg{c^{\dagger}_{r,\sigma}c_{r,\sigma}}-1}$ in the $N=L-2,S=0$ sector.
Lower right inset: spin density $\avg{S^z_r}=1/2\left(\avg{n_{r,\uparrow}}-\avg{n_{r,\downarrow}}\right)$ in the $N=L,S=1$ sector.
}
\end{figure}

\paragraph{Edge modes.—} 
With a diagonally cut edge (cf. Fig.~\ref{fig:ladder}), the repulsive Hubbard ladder ($U>0$, all $V$-terms equal zero) is known to host spin-1/2 edge modes protected by particle-hole symmetry~\cite{Sompet2022, VMP, White1996}.
We find that for our extended model, edge modes can carry either spin or charge quantum numbers, transforming differently under time-reversal symmetry. Intuitively, if on-site singlets (S-Mott) are cut, empty or doubly occupied sites with particle number $N=\pm2$ remain (see Fig.~\ref{fig:edgeStates}). A change of edge quantum numbers is induced when varying the interaction parameters $U$ and $V_{\perp}$.
From the model wave function, one might naively assume that the edge quantum number is directly related to the bulk being D-Mott or S-Mott (i.e., to the sign of $\langle O_j \rangle$), but this is not the case: We find that spinful edge states are strongly preferred, except for very small $U$. For example, for $V_{\perp}=5$, a change in quantum numbers already occurs at $U\approx 0.14$ (see Fig.~\ref{fig:edgeStates}), far away from the bulk crossover $U=V_{\perp}$.
Thus, our system provides an example where an edge transition  has no bulk indication~\cite{Abhi}; though further details are beyond the scope of this study.

\paragraph{Discussion.-}
We have shown that D-Mott and S- Mott are two facets of the same topological phase. An intuitive explanation is that true $d$-wave symmetry can only be found on the full 2D square lattice~\cite{Yao}. A robust terminable transition nevertheless exists without fine-tuning and can be understood with the help of the concept of singlet-density difference and an effective theory. 
Continuous transitions can be revealed experimentally by thermal conductivity peak with charge and spin transport less influenced.
The existence of a robust transition itself is assisted by time-reversal symmetry, which sheds light on the study of robust terminable transitions~\cite{Abhi}. 
We propose that our effective theory may be useful in discovering very different systems with similar unconventional transition behavior. 

\paragraph{Acknowledgments.---}
We thank  Fabian Essler, Sid Parameswaran, Abhishodh Prakash, and Limei Xu  for discussion. Y.H. and D.M.K. are supported by the Deutsche Forschungsgemeinschaft (DFG,
German Research Foundation) under RTG 1995, within the Priority Program
SPP 2244 ``2DMP" and under Germany's Excellence Strategy - Cluster of
Excellence Matter and Light for Quantum Computing (ML4Q) EXC 2004/1 -
390534769.  We acknowledge support by the Max Planck-New York City Center for Nonequilibrium Quantum Phenomena. The numerical calculations have been partially performed with computing resources granted by RWTH Aachen University under project rwth0726.

\bibliography{bib}

\clearpage

\section{Supplemental Material}
\renewcommand{\theequation}{S\arabic{equation}}
\setcounter{equation}{0}
\renewcommand{\thefigure}{S\arabic{figure}}
\setcounter{figure}{0}
\renewcommand{\thetable}{S\arabic{table}}
\setcounter{table}{0}

\subsection{Extrapolating correlation lengths and gaps}

To find the phase transition line where an excitation gap closes, we can compute the gap directly on finite systems of length $L$ and extrapolate to the infinite limit in $L^{-1}$. This can be resolved by quantum number: $\Delta_{\text{spin}} = E_0\lr{S=1,N=L}-E_0\lr{S=0,N=L}$ defines the spin gap, $\Delta_{\text{charge}} = E_0\lr{S=1/2,N=L+1}-E_0\lr{S=0,N=L}$ defines the charge gap, while $\Delta_{\text{neutral}} = E_1\lr{S=0,N=L}-E_0\lr{S=0,N=L}$ defines the neutral gap, whereby we label $E_0$ ($E_1$) the lowest (second lowest) eigenenergy in a given sector.

\begin{figure}
\begin{center}
\includegraphics[width=0.8\columnwidth]{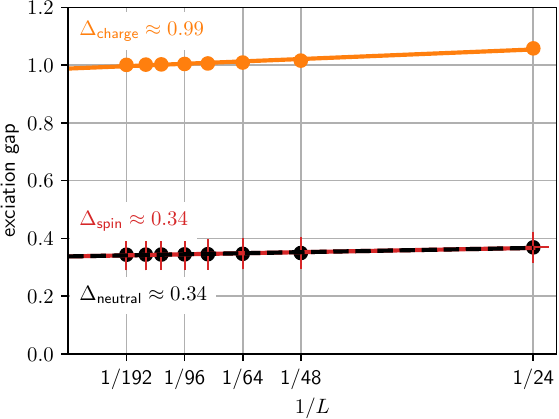}
\end{center}
\caption{\label{fig:UV4gap}
Gap extrapolation for $U=V_{\perp}=4$ (minimal model) for systems of length $L$ (with $L/2$ rungs). Energies are obtained using DMRG for finite ladders with a trivial cut (unlike Fig.5).
}
\end{figure}

The result of this process is shown in Fig.~\ref{fig:UV4gap} above the termination point, clearly showing that all gaps remain finite.

Another possibility is to compute the inverse correlation length $\xi^{-1}$ for the infinite system, which also goes to zero at the gap closure. The correlation length is in this case obtained from the dominant eigenvalue of the transfer matrix at a fixed bond dimension $\chi$~\cite{McCulloch2008,Rams_2018,VUMPS} and can also be resolved by the same quantum numbers as above. (Note that the main text shows the neutral correlation length measured in sites rather than unit cells).
One can use different extrapolation parameters $\delta$ that measure the closeness to the exact ground state (e.g. the inverse bond dimension $\chi^{-1}$). Here, we follow Ref.~\onlinecite{Rams_2018}, where a parameter was found with which $\xi^{-1}\lr{\delta}$ generally becomes linear.

\begin{figure}
\begin{center}
\includegraphics[width=0.8\columnwidth]{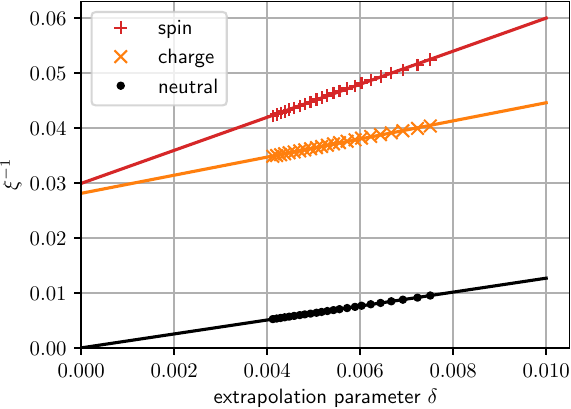}
\end{center}
\caption{\label{fig:xiExtrap_UV2}
Extrapolation of the inverse correlation length $\xi^{-1}$ (measured in sites and using the trivial cut) for $U=V_{\perp}=2$ (minimal model), resolved by the quantum number. The results were obtained using VUMPS for the infinite system. The extrapolation parameter $\delta$ is defined as in Ref.~\onlinecite{Rams_2018}.
}
\end{figure}

Figure~\ref{fig:xiExtrap_UV2} shows this procedure below the termination point. We find that the neutral correlation length vanishes, while charge and spin gaps remain open.

\subsection{Microscopic analysis for the role of the operator $O_j$}

In this section, we derive an understanding of the singlet density operator $O_j$ in terms of the subband picture of the ladder.

We repeat the Hamiltonian of the minimal model in the subband basis introduced in the main text:
\begin{align}
H=&-t_{\parallel}\sum_{j,k_y,\sigma}(c^{\dagger}_{j,k_y,\sigma}c_{j+1,k_y,\sigma}+h.c.)-t_{\perp}\sum_{j}[n_{j,\pi}-n_{j,0}]\nonumber \\
&+U/2 \sum_j [\Delta n_{j,\pi}+ \Delta n_{j,0}]^2-(U-V_{\perp})H_{\mathrm{res}}.
\end{align}
Rewriting $O_j$ in the same basis, we obtain:
\begin{align}\label{orderskbasis}
 O_j = -c^{\dagger}_{j,0,\uparrow}c^{\dagger}_{j,0,\downarrow} c_{j,\pi,\uparrow}c_{j,\pi,\downarrow}+H.c.
\end{align}
We see that $O_j$ describes the hopping of fermion pairs between the two subbands and can be used to characterize the strength of virtual scattering that violates the subband U(1) symmetry. Since the particle numbers in the subbands are conserved for $U=V_{\perp}$, the hopping between them must also vanish along this line: $\langle O_j \rangle=0$.

To understand that $\langle O_j \rangle$ and $U-V_{\perp}$ have the same sign, we note that the residual term is given by $H_{\mathrm{res}}=\sum_{j}O_j/2+...$, where the subband U(1) preserving terms have been neglected. Therefore, $-(U-V_{\perp})\langle O_j \rangle<0$ is expected to minimize the energy.

Similar to the above energy minimization argument for $\langle O_j \rangle$, we can argue that if there are subband U(1) violating terms breaking time-reversal symmetry (TRS) $\propto \sum_j e^{i\alpha}c^{\dagger}_{j,0,\uparrow}c^{\dagger}_{j,0,\downarrow} c_{j,\pi,\uparrow}c_{j,\pi,\downarrow}+H.c.$, the subband U(1) point can be avoided along a path connecting D-Mott to S-Mott by tuning $\alpha$ from 0 to $\pi$. This indicates that TRS plays an important role for the existence of a transition. This analysis is closely related  to the bosonization analysis of discreetness of locking values in the main text.

As shown in the main text, the ground state $\avg{n_{j,\pi}}=0$ and $\avg{n_{j,0}}=2$ for $U=V_{\perp}\gtrsim3.4$ can be written as rung bisinglet $\prod_{j} 1/\sqrt{2}(\Delta^{\dagger}_{\mathrm{S}j}+\Delta^{\dagger}_{\mathrm{D}j})|\Omega\rangle$.  Recall from the main text that the model wave function of D- and S-Mott ($|\mathrm{D}\rangle=\prod_{j} \Delta^{\dagger}_{\mathrm{D}j}|\Omega\rangle$ and  $|\mathrm{S}\rangle=\prod_{j} \Delta^{\dagger}_{\mathrm{S}j} |\Omega\rangle$) consists rungs of two eigenstates of $O$ operators. Consider $O$ as an Ising field, the tendency to form rung bisinglet can be induced by transverse field, which is the counter part of the temperature in the quantum-classical  analogy. This is one way to draw an analogy to the magnet picture of terminable first-order transition. However, with accidental symmetry or weak interaction, the analogues of magnetic ordering terms can vanish simultaneously (see the scaling dimension analysis in the main text) such that the transition is not first order as the magnetic picture.

\subsection{Effective band structure}

In this section, we offer a perspective on the termination of the phase transition for the minimal model from the effective band structure.

The criticality at $U=V_{\perp}$ and its termination is related to the subband occupation ratio $\avg{n_{j,\pi}}/\avg{n_{j,0}}$ (which is not dependent on $j$ in the homogeneous case).  When this filling ratio is fractional, according to Lieb-Schultz-Mattis theorem~\cite{LSM,Tasaki2018}, the system must be gapless, unless there is a spontaneous breaking of translational symmetry resulting in a degenerate ground state.

Introducing the single-particle retarded Green's function
\begin{equation}\label{eq:Gt}
\begin{split}
G^{ll'}\lr{t,\big|j-j'\big|} = &-i\theta\lr{t}  
\big[
\sum_{\sigma}\matrixel{0}{e^{iHt}c^{\dagger}_{jl\sigma}e^{-iHt}c_{j'l'\sigma}}{0} \\
&+\sum_{\sigma}\matrixel{0}{e^{-iHt}c_{jl\sigma}e^{iHt}c^{\dagger}_{j'l'\sigma}}{0} 
\big]
\end{split}
\end{equation}
and its Fourier transform
\begin{equation}
G^{ll'}\lr{\omega,k} = \sum_d e^{ikd} \int_{-\infty}^{\infty} dt~ e^{i\omega t} G^{ll'}(t,d),
\end{equation}
we can define the equivalent of the bandstructure in presence of interactions by the spectral function
\begin{equation}\label{eq:S_omega_k}
S\lr{\omega,k} = -\frac{1}{\pi} \sum_{l=\text{A},\text{B}} \text{Im}~ G^{ll}(\omega,k).
\end{equation}

\begin{figure}
\includegraphics[width=\columnwidth]{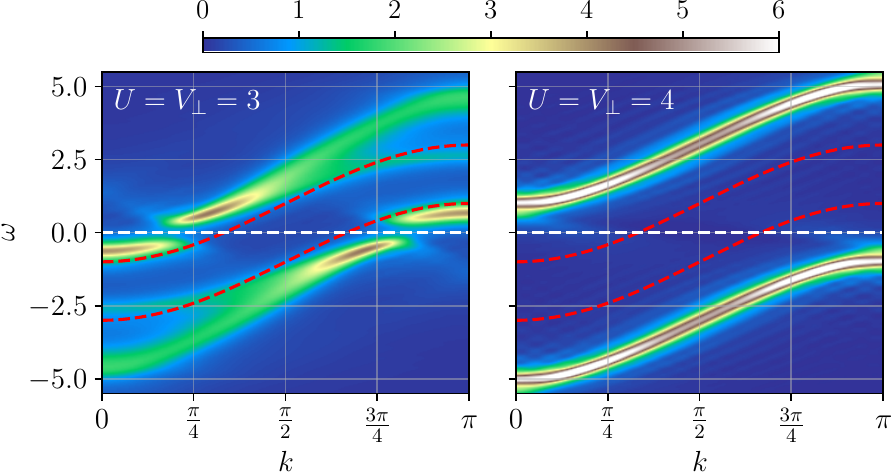}
\caption{
\label{fig:kspec}
The spectral function Eq.~\eqref{eq:S_omega_k}. Red dashed line: band
 structure of the noninteracting model; white dashed line: reference for $\omega=0$. The results are obtained by a 
real-time evolution of Eq.~\eqref{eq:Gt} to $t_{\text{max}}=20$ using infinite boundary conditions~\cite{Phien_2012,Haegeman_2016}.
}
\end{figure}

This spectral function is displayed in Fig.~\ref{fig:kspec}. It reveals the two-subband structure of the ladder, whereby the lower subband has $k_y=0$ and the upper subband has $k_y=\pi$.
The parts of the subbands that lie below the Fermi edge $\omega=0$ reflect $\avg{n_{j,\pi}}$ and $\avg{n_{j,0}}$ when integrated.
In the noninteracting limit we have $\avg{n_{j,\pi}}/\avg{n_{j,0}}=1/2$. Our calculations show that $\avg{n_{j,\pi}}/\avg{n_{j,0}}$ can change continuously along the line $U=V_{\perp}$. The effect of interactions is to increase the splitting of the subbands, so that for $U=V_{\perp}\gtrsim3.4$, only the lower band is below the Fermi energy. This implies an integer filling $\avg{n_{j,\pi}}=0$ and $\avg{n_{j,0}}=2$ and the state effectively becomes a band insulator, where the two-band bosonization is no longer valid. This is why within the two-band bosonization, it is not clear that the two Mott regions can be adiabatically connected.

%
%

\subsection{The bosonization of $O_j$ }

Here, we discuss the two-band bosonization of $O_j$. Recall that the definition is 
\begin{align}\label{suporderskbasis}
 O_j=-c^{\dagger}_{j,0,\uparrow}c^{\dagger}_{j,0,\downarrow} c_{j,\pi,\uparrow}c_{j,\pi,\downarrow}+H.c.
\end{align}
In addition to the terms presented in the main text, we include oscillatory terms and discuss higher harmonics. Recall that the bosonization of fermion operators to the lowest harmonics is 
\begin{align}\label{bosonization}
c_{k_y,\sigma}(x_j)
=
\frac{\kappa_{k_y,\sigma}}{\sqrt{2\pi}} \sum_{\eta=-1,1}e^{i[\theta_{k_y,\sigma}+\eta(\phi_{k_y,\sigma}+k_{\text{F},k_y,\sigma}x_j)]}.
\end{align}

We insert Eq.~\eqref{bosonization} into Eq.~\eqref{suporderskbasis} to obtain the lowest harmonics of the bosonization of  $O_{j}$
\begin{align}\label{Obosonization}
O_{j} \propto & \cos(2\tilde{\theta}_{c,-})[\cos(2\tilde{\phi}_{s,-})+\cos(2\tilde{\phi}_{s,+}) + \nonumber \\ 
& \cos(2\tilde{\phi}_{c,+}+\sum_{\sigma,k_y}k_{\mathrm{F},k_y,\sigma}x_j)+ \nonumber \\
& \cos(2\tilde{\phi}_{c,-}+\sum_{\sigma}(k_{\mathrm{F},\pi,\sigma}-k_{\mathrm{F},0,\sigma})x_j)+ \nonumber \\
& \cos(\tilde{\phi}_{c,+}+\tilde{\phi}_{c,-}+\tilde{\phi}_{s,+}-\tilde{\phi}_{s,-}+\sum_{\sigma}k_{\mathrm{F},0,\sigma} x_j) +\nonumber \\
&\cos(\tilde{\phi}_{c,+}-\tilde{\phi}_{c,-}+\tilde{\phi}_{s,+}+\tilde{\phi}_{s,-}+\sum_{\sigma}k_{\mathrm{F},\pi,\sigma} x_j)+..., 
\end{align}
where the higher harmonics are neglected.
The coefficients of each term are neglected for simplicity. 
At half filling, $\sum_{\sigma,k_y}k_{\mathrm{F},k_y,\sigma}=2\pi$, which can be set to be 0, as $x_j$ are integer.
For the Mott states, as well as the D-Mott/S-Mott transition (shown to be of Gaussian type, see below), $\tilde{\phi}_{c,+}$, $\tilde{\phi}_{s,-}$,$\tilde{\phi}_{s,+}$ are kept locked and can be set to zero for discussing expectation values or correlation function. So in this special case, we have 
\begin{align}\label{Obosonizations}
O_{j} \propto & \cos(2\tilde{\theta}_{c,-})+\cos(2\tilde{\theta}_{c,-})[\cos(2\tilde{\phi}_{c,-}+2\Delta k_{\mathrm{F}} x_j)+\nonumber\\ &\cos(\tilde{\phi}_{c,-}+2k_{\mathrm{F},0,\sigma} x_j)+\cos(\tilde{\phi}_{c,-}+2k_{\mathrm{F},\pi,\sigma} x_j)],
\end{align}
where $\Delta k_{\mathrm{F}}=k_{\mathrm{F},0,\sigma}-k_{\mathrm{F},\pi,\sigma}$; for our model, $k_{\mathrm{F},k_y,\uparrow}=k_{\mathrm{F},k_y,\downarrow}$. The coefficient of each term is neglected for simplicity. These are  only two values $\tilde{\theta}_{c,-}$ can be locked at if there is no explicit or spontaneous TRS breaking; this can be seen by evaluating TRS odd term $i(c^{\dagger}_{j,0,\uparrow}c^{\dagger}_{j,0,\downarrow} c_{j,\pi,\uparrow}c_{j,\pi,\downarrow}-h.c) \propto \sin(2\tilde{\theta}_{c,-})$. In fact, $\tilde{\theta}_{c,-} \rightarrow -\tilde{\theta}_{c,-}$ for time-reversal symmetry, thus $\cos(2\tilde{\theta}_{c,-}+\alpha)$ with generic $\alpha$ is forbidden to appear in a time-reversal symmetric Hamiltonian. The locking values are 0 and $\pi/2$ Thus, with TRS, the expected continuous transition within the 2-band effective theory is that $\tilde{\theta}_{c,-}$ becomes unlocked, indicating a Gaussian criticality. This is consistent with the microscopic argument of the TRS's role for the existence of the transition.

\subsection{Correlations $\langle O_j O_{j+d}\rangle$ and Gaussian criticality}

\begin{figure}
\begin{center}
\includegraphics[width=\columnwidth]{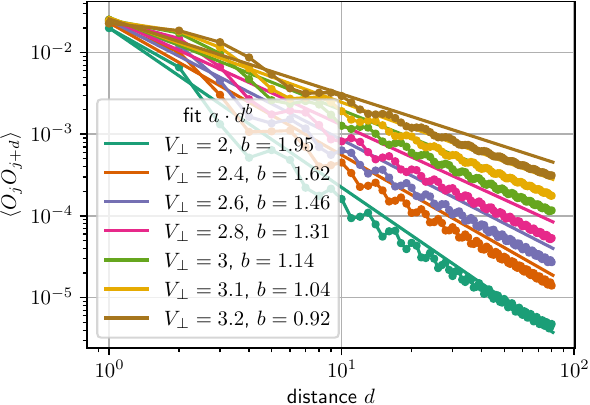}
\end{center}
\caption{Correlation function $ \langle O_j O_{j+d}\rangle$ for $U=V_{\perp}$ for the minimal model obtained in the infinite system. The straight solid lines are references for the fitted slope of the data with the same color.
\label{fig:OO}
}
\end{figure}

For the minimal model and $U=V_{\perp}\lesssim 3.4$, as pointed out in the main text, the field $\tilde{\theta}_{c,-}$. and its dual field $\tilde{\phi}_{c,-}$ is gapless and characterized by a Luttinger parameter $K$ (Eq.~(8)). Here we compute $\langle O_j O_{j+d}\rangle$ using Eq.~\eqref{Obosonizations} of which each term gives an algebraic decay component.
\begin{align}\label{OOterms}
\langle O_j O_{j+d}\rangle=&\frac{1}{|d|^{2/K}}+\frac{\cos(2k_{\mathrm{F},0,\sigma} d)}{|d|^{2/K+K/2}}+\frac{\cos(2k_{\mathrm{F},\pi,\sigma} d)}{|d|^{2/K+K/2}}+ \nonumber\\
&\frac{\cos(2\Delta k_{\mathrm{F}} d)}{|d|^{2/K+2K}}
\end{align}
The coefficient of each term is neglected for simplicity. We see that the leading term is non-oscillatory while subleading terms can be oscillatory.

The data of $\langle O_j O_{j+d}\rangle$ for various $U=V_{\perp}$ are plotted in Fig.~\ref{fig:OO}. We fit the exponent $2/K$ of the leading term using log-log scale data. The fitted result for $2/K$ decreases from $\sim 0.96$ to $\sim 0.46$ as $U=V_{\perp}$ is increased from $2$ to $3.2$. From Fig.~\ref{fig:OO}, we also observe subleading oscillations. In our predictions Eq.~\eqref{fig:OO}, these subleading exponents are at least larger than the leading exponent by addition of $K/2$. Using the fitting result of $2/K$, this indicates that the exponent difference $K/2$ ranges from $\sim 1.04$ to $\sim 2.17$. Observing nonuniversal exponents numerically, we conclude that the transition line is of Gaussian type.

\subsection{Corrected bosonization of $\Delta_{\mathrm{D}}$ and $\Delta_{\mathrm{S}}$ }

The two-band bosonization previously reported in the literature has missed the possibility of a terminated transition. In this section, we present a corrected way of doing two-band bosonization of $\Delta_{\mathrm{D}}$ and $\Delta_{\mathrm{S}}$. Recall from Eq.~(5),
\begin{align}\label{eq2r}
& \Delta_{\mathrm{D}j}=(c_{j,\mathrm{A},\uparrow} c_{j,\mathrm{B},\downarrow}+c_{j,\mathrm{B},\uparrow} c_{j,\mathrm{A},\downarrow})/\sqrt{2}, \nonumber \\
& \Delta_{\mathrm{S}j}=(c_{j,\mathrm{A},\uparrow} c_{j,\mathrm{A},\downarrow}+c_{j,\mathrm{B},\uparrow} c_{j,\mathrm{B},\downarrow})/\sqrt{2}. 
\end{align}
We note that $\Delta_{\mathrm{D}}$ and $\Delta_{\mathrm{S}}$ do not transform with different parity under any symmetry of the Hamiltonian. We aim to reconcile this fact with the fact that using bosonization, these two order parameters appear to give separate quasi-long-range orders in $d$- and $s$- paired liquids respectively. (The $d$-wave and $s$-wave states are the doped D-Mott and S-Mott respectively.) With the correction, bosonization also concludes that their existence is not mutually exclusive. We will discuss the microscopic definition of $s$- and $d$- paired liquids.

We introduce the symbols
\begin{align}\label{lrmovers}
\psi_{k_y,\eta,\sigma}=\frac{\kappa_{k_y,\sigma}}{\sqrt{2\pi}} e^{i(\theta_{k_y,\sigma}+\eta\phi_{k_y,\sigma})}
\end{align}
for convenience. Then we can write Eq.~\eqref{bosonization} as 
\begin{align}
c_{k_y,\sigma}(x_j)
= \sum_{\eta=-1,1}e^{ik_{\text{F},k_y\sigma}x_j}\psi_{k_y,\eta,\sigma}+...,
\end{align}
where ... represents neglected higher harmonics.
\begin{align}
&\Delta_{\mathrm{S}}=\sum_{\eta,\eta'=\pm 1} [\psi_{0,\eta,\uparrow} \psi_{0,\eta',\downarrow}+\psi_{\pi,\eta,\uparrow} \psi_{\pi,\eta',\downarrow}]+..., \nonumber \\  
&\Delta_{\mathrm{D}}=\sum_{\eta,\eta'=\pm 1} [\psi_{0,\eta,\uparrow} \psi_{0,\eta',\downarrow}-\psi_{\pi,\eta,\uparrow} \psi_{\pi,\eta',\downarrow}]+...,  
\end{align}
where higher harmonic terms have been neglected. Using Eq.~\eqref{lrmovers}, we obtain
\begin{align}\label{kypairing}
&\sum_{\eta,\eta'=\pm 1} \psi_{0,\eta,\uparrow} \psi_{0,\eta',\downarrow} \nonumber \\ 
&= C_0 e^{i\sum_{\sigma}\theta_{0,\sigma}}\cos(\phi_{0,\uparrow}-\phi_{0,\downarrow})+...\nonumber\\
&=C_0 e^{i\sum_{\sigma}\theta_{0,\sigma}}[\cos(\tilde{\phi}_{+,s})\cos(\tilde{\phi}_{-,s})-\sin(\tilde{\phi}_{+,s})\sin(\tilde{\phi}_{-,s})]+...\nonumber \\
&\sum_{\eta,\eta'=\pm 1}  \psi_{\pi,\eta,\uparrow} \psi_{\pi,\eta',\downarrow} \nonumber \\
&=C_{\pi}  e^{i\sum_{\sigma}\theta_{\pi,\sigma}}\cos(\phi_{\pi,\uparrow}-\phi_{\pi,\downarrow})+... \nonumber\\
&=C_{\pi} e^{i\sum_{\sigma}\theta_{\pi,\sigma}}[\cos(\tilde{\phi}_{+,s})\cos(\tilde{\phi}_{-,s})+\sin(\tilde{\phi}_{+,s})\sin(\tilde{\phi}_{-,s})]+...
\end{align}
where oscillatory terms and higher harmonics have been neglected; we explicitly write out the coefficients $C_0$ and $C_{\pi}$, which depend on parameters of Hamiltonian. The subtle issue is that with interaction, one cannot correctly obtain the values of $C_0$ and $C_{\pi}$ by a naive multiplication of vertex operators' coefficients. Unlike the noninteracting limit where $C_0=C_{\pi}$ because there is no exchange symmetry of $\pi$ and $0$ bands, we expect $C_0 \neq C_{\pi}$ for general interacting cases. Using the convention that $\tilde{\phi}_{-,s}$ and $\tilde{\phi}_{-,s}$ are locked at 0 for $d$- and $s$- wave paring and for the purpose of evaluating their quasi-long-range orders, Eq.~\eqref{eq2r} can be written as  
\begin{align}
&\Delta_{\mathrm{S}} \propto e^{i\tilde{\theta}_{+,c}}[C_0\cos(\tilde{\theta}_{-,c})+(C_{\pi}-C_0)e^{i\tilde{\theta}_{-,c}}] +... \nonumber \\  
&\Delta_{\mathrm{D}}\propto e^{i\tilde{\theta}_{+,c}}[C_0\sin(\tilde{\theta}_{-,c})-(C_{\pi}-C_0)e^{i\tilde{\theta}_{-,c}}]+...,  
\end{align}
In the convention, $\tilde{\theta}_{-,c}$ is locked at 0 and $\pi/2$ for $s$- and $d$- pairing states  respectively. Thus if $C_0=C_\pi$, only quasi-long-range order of  $\Delta_{\mathrm{S}}$ exists for $s$-wave states; the same applies to $d$-wave. Given that $C_0 \neq C_\pi$ in general, we have both quasi-long-range orders in either $s$- and $d$-wave states; in other words, $\langle \Delta_{\mathrm{S}j} \Delta^{\dagger}_{\mathrm{S},j+d} \rangle$ and $\langle \Delta_{\mathrm{D}j} \Delta^{\dagger}_{\mathrm{D},j+d} \rangle $ decay algebraically in $|d|$ with the same exponent. However, we can still have a microscopic definition of $s$-wave and $d$-wave states. If the leading algebraic decay prefactor of $\langle \Delta_{\mathrm{S}j} \Delta^{\dagger}_{\mathrm{S},j+d} \rangle$ is larger, we call the state $s$-wave, otherwise $d$-wave. This is equivalent to the definition from the relative sign of the coefficient of the leading  algebraic decay components of  $\langle \Delta_{\mathrm{0,j}} \Delta^{\dagger}_{\mathrm{0,j+d}} \rangle$ and $\langle \Delta_{\mathrm{\pi,j}} \Delta^{\dagger}_{\mathrm{\pi,j+d}} \rangle$, where $\Delta_{\mathrm{k_y,j}}=c_{j,k_y,\uparrow}c_{j,k_y,\downarrow}$ A positive relative sign is defined as $s$-wave states and a negative relative sign is defined as $d$-wave states.
Such a definition no longer necessitates a transition between $s$- and $d$-states. A definition via the sign is closely related to our definition of S- and D-Mott using the sign of $\langle O_j \rangle$.

\subsection{Analytical solution for a single rung}
In this section, we discuss the analytical solution of a single rung at half filling (which amounts to analyzing a 4$\times$4 matrix) and discuss what one can learn from it for the full model.

The Hamiltonian of the minimal model on a single rung can be written as:
\begin{equation}
\begin{split}
H_{\text{rung}} &= 
-t_{\perp} \sum_{\sigma} \left( c^{\dagger}_{\mathrm{A}\sigma}c_{\mathrm{B}\sigma} + h.c. \right)\\
&+U\left(n_{\mathrm{A}\uparrow}n_{\mathrm{A}\downarrow}+n_{\mathrm{B}\uparrow}n_{\mathrm{B}\downarrow} \right)\\
&+V_{\perp}\sum_{\sigma\sigma'} n_{\mathrm{A}\sigma} n_{\mathrm{B}\sigma'}
+V_{\perp}
-V_{\perp}\sum_{\sigma}\left(n_{\mathrm{A}\sigma}+n_{\mathrm{B}\sigma}\right).
\end{split}
\end{equation}
At half filling we can replace $\sum_{\sigma}\lr{n_{\mathrm{A}\sigma}+n_{\mathrm{B}\sigma}}=2$ and use the basis states $c^{\dagger}_{\mathrm{A}\uparrow}c^{\dagger}_{\mathrm{B}\downarrow}\ket{\Omega}=\ket{\uparrow,\downarrow}$, $c^{\dagger}_{\mathrm{A}\downarrow}c^{\dagger}_{\mathrm{B}\uparrow}\ket{\Omega}=\ket{\downarrow,\uparrow}$, $c^{\dagger}_{\mathrm{A}\uparrow}c^{\dagger}_{\mathrm{A}\downarrow}\ket{\Omega}=\ket{\uparrow\downarrow,0}$, $c^{\dagger}_{\mathrm{B}\downarrow}c^{\dagger}_{\mathrm{B}\uparrow}\ket{\Omega}=\ket{0,\uparrow\downarrow}$, whereby $\ket{\Omega}$ is the vacuum. In this basis, the Hamiltonian matrix reads:
\begin{equation}
H=
\begin{pmatrix}
0 & 0 & -t_{\perp} & -t_{\perp}\\
0 & 0 & +t_{\perp} & +t_{\perp}\\
-t_{\perp} & +t_{\perp} & U-V_{\perp} & 0\\
-t_{\perp} & +t_{\perp} & 0 & U-V_{\perp}
\end{pmatrix}.
\end{equation}

This is a two-site Hubbard problem, extended by $V_{\perp}$. We see that for a single rung, its effect is to simply shift $U\to U-V_{\perp}$ (which is not generally true for the full ladder). The eigenstates can be characterized by the spin and pseudospin quantum numbers (which we call ``S'' and ``T'', respectively), whereby the pseudospin operators for a bipartite lattice are in general defined as:
\begin{equation}
\begin{split}
T_i^{-} &= (-1)^i c_{i\downarrow}c_{i\uparrow}, \\
T_i^{+} &= (-1)^i c^{\dagger}_{i\uparrow}c^{\dagger}_{i\downarrow},\\
T_i ^z &= \frac{1}{2} \left(n_i-1\right),
\end{split}
\end{equation}
and fulfill SU(2) algebra relations $\left[T^z_i,T^{\pm}_j\right]=\pm \delta_{ij} T^{\pm}_i$, $\left[T^{+}_i,T^{-}_j\right]=2\delta_{ij}T^z_i$. For a single rung the indices are $i,j=\text{A},\text{B}$.

Two of the four eigenstates are the spin-triplet and the pseudospin-triplet:
\begin{itemize}
\item spin-triplet:\\
$\ket{S=1,M_S=0} = \frac{1}{\sqrt{2}} \left( \ket{\uparrow,\downarrow} + \ket{\downarrow,\uparrow} \right)$\\
$E=0$
\item pseudospin-triplet:\\
$\ket{T=1,M_T=0} = \frac{1}{\sqrt{2}} \left( \ket{0,\uparrow\downarrow} - \ket{\uparrow\downarrow,0} \right)$\\
$E=U-V_{\perp}$
\end{itemize}

The corresponding singlets are:
\begin{itemize}
\item spin-singlet:\\
$\ket{S=0} = \frac{1}{\sqrt{2}} \lr{ \ket{\uparrow,\downarrow} - \ket{\downarrow,\uparrow} }$
\item pseudospin-singlet:\\
$\ket{T=0} = \frac{1}{\sqrt{2}} \lr{ \ket{0,\uparrow\downarrow} + \ket{\uparrow\downarrow,0} }$
\end{itemize}
However, they are not by themselves eigenstates. Instead, one needs to form a ``bonding'' and and an ``antibonding'' superposition:
\begin{itemize}
\item bonding singlet superposition:\\
$\ket{S=T=0,-} = \alpha_-\ket{S=0} + \beta_-\ket{T=0}$\\
$E_-=\frac{U-V_{\perp}}{2}-\sqrt{\lr{\frac{U-V_{\perp}}{2}}^2+4t_{\perp}^2}$
\item antibonding singlet superposition:\\
$\ket{S=T=0,+} = \alpha_+ \ket{S=0} + \beta_+\ket{T=0}$\\
$E_+=\frac{U-V_{\perp}}{2}+\sqrt{\lr{\frac{U-V_{\perp}}{2}}^2+4t_{\perp}^2}$
\end{itemize}
The mixing coefficients are given by:
\begin{equation}
\begin{split}
\alpha_{\pm} &= \frac{1}{\sqrt{1+E_{\pm}^2/(4t_{\perp}^2)}}, \\
\beta_{\pm} &= -\frac{E_{\pm}}{2t_{\perp}} \frac{1}{\sqrt{1+E_{\pm}^2/(4t_{\perp}^2)}}.
\end{split}
\end{equation}

The bonding singlet superposition is always the ground state. It contains more spin-singlets in the admixture for $U>V_{\perp}$ (which becomes the D-Mott phase on the ladder), more ``pseudo-singlets'' (on-site singlets) for $U<V_{\perp}$ (which becomes the S-Mott phase); and an equal superposition $\frac{1}{\sqrt{2}} \left(\ket{S=0}+\ket{T=0}\right)$ for $U=V_{\perp}$, which we call a ``rung bisinglet'' in the main text.

In the strong-coupling limit $\big|U-V_{\perp}\big| \gg t_{\perp}$ and for $U>V_{\perp}$, we have $E_- \approx -4\frac{t_{\perp}^2}{U-V_{\perp}}=J$, and the two low-lying states become the spin-singlet and spin-triplet, split in energy by $J$, indicating an effective Heisenberg model.

On the other hand, if $V_{\perp}<U$, we obtain $E_- \approx U-V_{\perp}+J$ and the two low-lying states become the pseudospin singlet and pseudospin triplet, again split in energy by $J$. We see that even though the density-density interaction of the original model is of Ising type and only couples the z-components of the pseudospin, the strong-coupling limit favors entangled singlet states. This is because we have restricted ourselves to half filling for the rung, where $V_{\perp}$ acts exactly as an attractive $U<0$.

For the full ladder, both $V_{\perp}>0$ and $U<0$ favor an S-Mott phase, but the effect of $V_{\perp}$ cannot be simply captured by substituting $U\to U-V_{\perp}$. Doing so neglects charge fluctuations on the rungs and will not reveal the terminated transition.

\end{document}